\documentclass[
 reprint,
superscriptaddress,https://v2.overleaf.com/project/5d362e0d09d8187cf105af8e
showpacs,
amsmath,amssymb,
 aps,prl,
]{revtex4-1}

\usepackage{graphicx}
\usepackage{dcolumn}
\usepackage{bm}
\usepackage{amsmath}
\usepackage{natbib}

\begin{document}

\preprint{APS/123-QED}

\title{Using Acoustic Perturbations to Dynamically Tune Shear Thickening in Colloidal Suspensions}

\author{Prateek Sehgal*}
\affiliation{Sibley School of Mechanical and Aerospace Engineering, Cornell University, Ithaca, New York 14853, USA}
\author{Meera Ramaswamy*}
\affiliation{Department of Physics, Cornell University, Ithaca, New York 14853, USA}
\author{Itai Cohen}
\affiliation{Department of Physics, Cornell University, Ithaca, New York 14853, USA}
\altaffiliation[Also at ]{Department of Medicine, Division of Hematology and Medical Oncology, Weill--Cornell Medicine, New York, New York 10021,USA}
\author{Brian J. Kirby}
\affiliation{Sibley School of Mechanical and Aerospace Engineering, Cornell University, Ithaca, New York 14853, USA}
\affiliation{Department of Medicine, Division of Hematology and Medical Oncology, Weill--Cornell Medicine, New York, New York 10021,USA}

\date{\today}

\begin{abstract}
Colloidal suspensions in industrial processes often exhibit shear thickening that is difficult to control actively. Here, we use piezoelectric transducers to apply acoustic perturbations to dynamically tune the suspension viscosity in the shear-thickening regime. We attribute the mechanism of dethickening to the disruption of shear-induced force chains via perturbations that are large relative to the particle roughness scale. The ease with which this technique can be adapted to various flow geometries makes it a powerful tool for actively controlling suspension flow properties and investigating system dynamics. 
\end{abstract}

\pacs{Valid PACS appear here}

\maketitle

The orders-of-magnitude increase in viscosity that arises under high shear makes dense suspensions ideal for numerous industrial applications including shock absorption, damping, soft-body armor, astronaut suits, and curved-surface polishing  \cite{BrownandJaeger2014,Wagner_PhysTod,Fischer_damping, Wagner_KevlarArmor, Li_STFPolishing,Wagner_Micrometeroid}. The challenge in using such shear thickening fluids, however, is that this same increase in viscosity can lead to jamming and failure of pumping and mixing equipment driving the flows. The ability to manage these limitations of this important technological material remains challenging \cite{BrownandJaeger2014, Barnes_review} because it requires actively tuning the suspension viscosity. Shear thickening viscosity previously has been tuned passively by changing the physical properties of the suspension constituents, such as the volume fraction ($\phi$) \cite{Cwalina_VolumeFraction,MariandSeto_ForceChains, Denn_Review}, particle size \cite{Guy_Friction}, particle shape \cite{Cwalina_Shape, hsiao2017rheological}, roughness \cite{Hsu_Tribology,Lootens_Roughness}, surface chemistry \cite{Yang_FrictionCST}, and solvent attributes \cite{Wagner_PhysTod, Brown2010, james2018interparticle, maranzano2001effects}; all of which affect the formation of the force chains responsible for thickening \cite{Cates_ForceChains, Yang_FrictionCST, Majmudar_ForceChains, MariandSeto_ForceChains,SetoandMari_FrictionDST, Morris_Review,Bi_jamming, Comtet_fricprofile, Clavaud_FrictionTransition,Lin_ShearReversal,Royer_Friction, Fernandez_FricTransition, SetoandMari_FrictionDST, Wyart_DSTFriction, BrownJaeger2012, Guy_Friction, Thomas_FrictionDST, rathee2017localized}. However, active tuning to change the flow properties on demand without changing the physical properties of the suspension constituents or without modifying the suspension has until recently remained largely unexplored.

Recently, it was shown that macroscopic boundary oscillations can be used to actively tune shear thickening in a dense suspension \cite{Lin_dethick}. The dethickening mechanism entails disruption of the force chains through application of an oscillatory shear flow orthogonal to the primary flow direction. Further simulations explored the parameter space for active tuning and showed that this mechanism is robust and that such orthogonal mechanical perturbations can tune the suspension viscosity over a wide range of shear rates and volume fractions \cite{Ness_CrossShear}. Unfortunately, using macroscopic boundary oscillations to introduce orthogonal perturbations is not practical for many applications. 

\begin{figure}[b]
\includegraphics[width=0.85\linewidth]{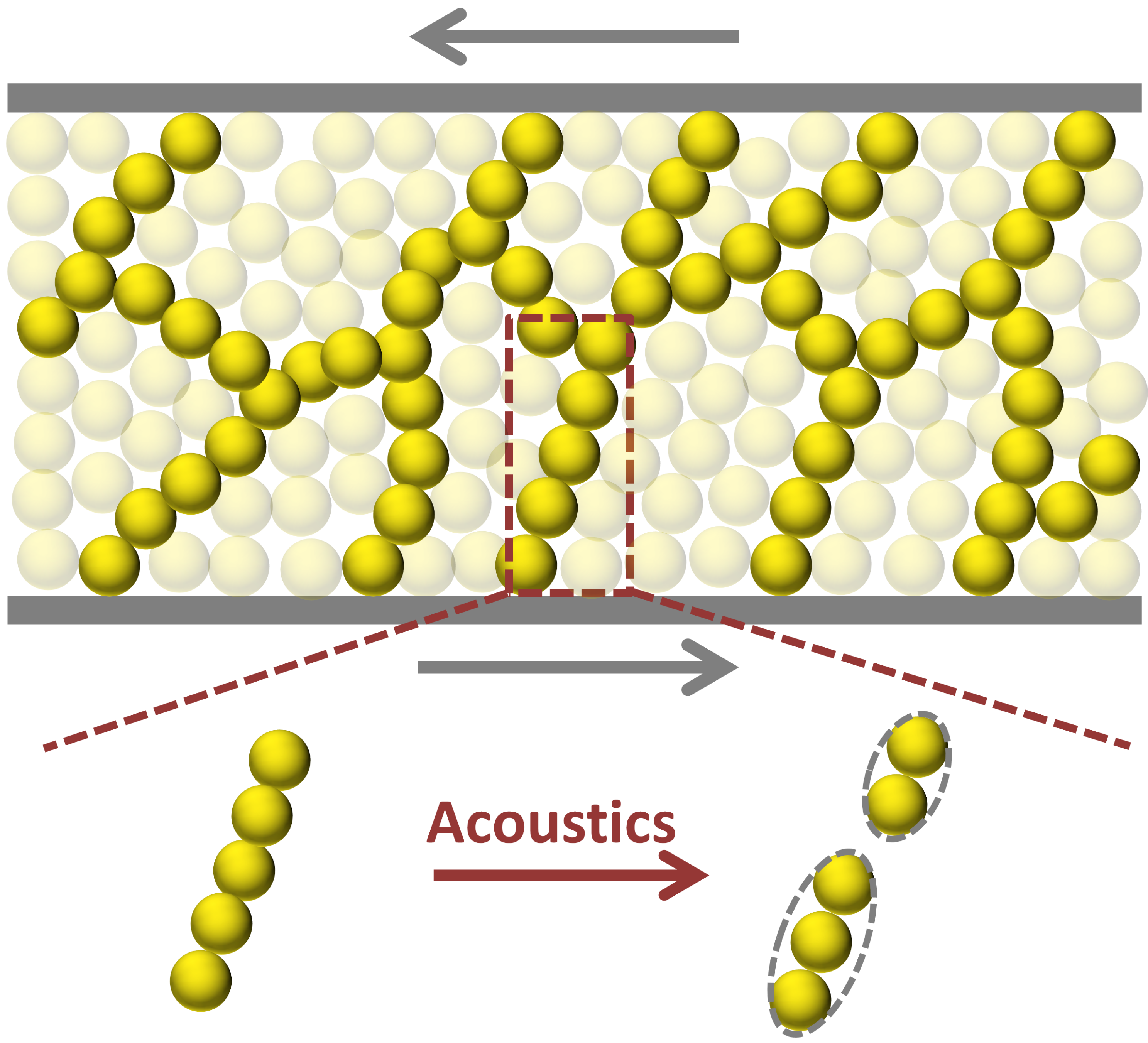}
\caption{\label{fig1} Hypothesized mechanism of dethickening. (Top) A schematic of the force chain network that forms in dense colloidal suspensions under shear. Grey arrows indicate the shear direction. (Bottom) Spatially non-uniform displacements of the particles in an acoustic field break the fragile force chains, and reduce the viscosity.}
\end{figure}

Here, we determine whether externally applied acoustic perturbations can be used to actively tune the suspension viscosity in the shear thickening regime. The advantage of this approach is that acoustic perturbations can controllably manipulate particles \cite{Collins_AcousticParticles, Sehgal_ParticleManipulation, Naseer_ParticlePatterning, Jaeger_ParticleAcoustics,Antfolk_SubmicronAcoustics, Destgeer_MicroparticleAssembly,Muller_3DParticleMotion,Ding_AcousticParticleManipulation} and can be applied via readily available piezoelectric transducers that are bonded to otherwise fixed surfaces \cite{Antfolk_SubmicronAcoustics, Laurell_AcousticManipulation, Lenshof_AcousticResonators}.
The key principle motivating our work is that nanoscale acoustic disturbances will locally perturb particles and break the force chains responsible for thickening (Fig. 1). From previous studies \cite{james2018interparticle, hsiao2017rheological, denn2014rheology, Lin_dethick, Hsu_Tribology}, we hypothesize that, to break force chains and achieve dethickening, such perturbations should generate relative particle displacements greater than the particle surface roughness scale and on time scales faster than the turnover rate for the force chains. In our system, the acoustic perturbations can generate non-uniform particle displacements \cite{Bruus_Acoustofluidics7} over the length of the force chains, and because these force chains are fragile \cite{Cates_ForceChains, Majmudar_ForceChains}, there can be a weak link in the chain where two particles are displaced beyond the surface roughness, resulting in breaking of the chain. In addition, in dense suspensions where interparticle distance is small, the particle contacts may further be disrupted by the acoustic scattering effects from nearby particles \cite{SilvaAndBruus2014} and bulk flows due to acoustic dissipation \cite{Wiklund_Streaming}. Furthermore, a typical ultrasonic transducer can apply these acoustic perturbations at time scales that is orders of magnitude shorter than the force chain turnover time, which is typically $\sim 1/\dot{\gamma}$ and O($\sim$1s) in our system, where $\dot{\gamma}$ is the strain rate. Thus, we anticipate that acoustic perturbations can disrupt the force chains in a thickened suspension and actively tune the viscosity.

\begin{figure}[t]
\includegraphics[width=0.8\linewidth]{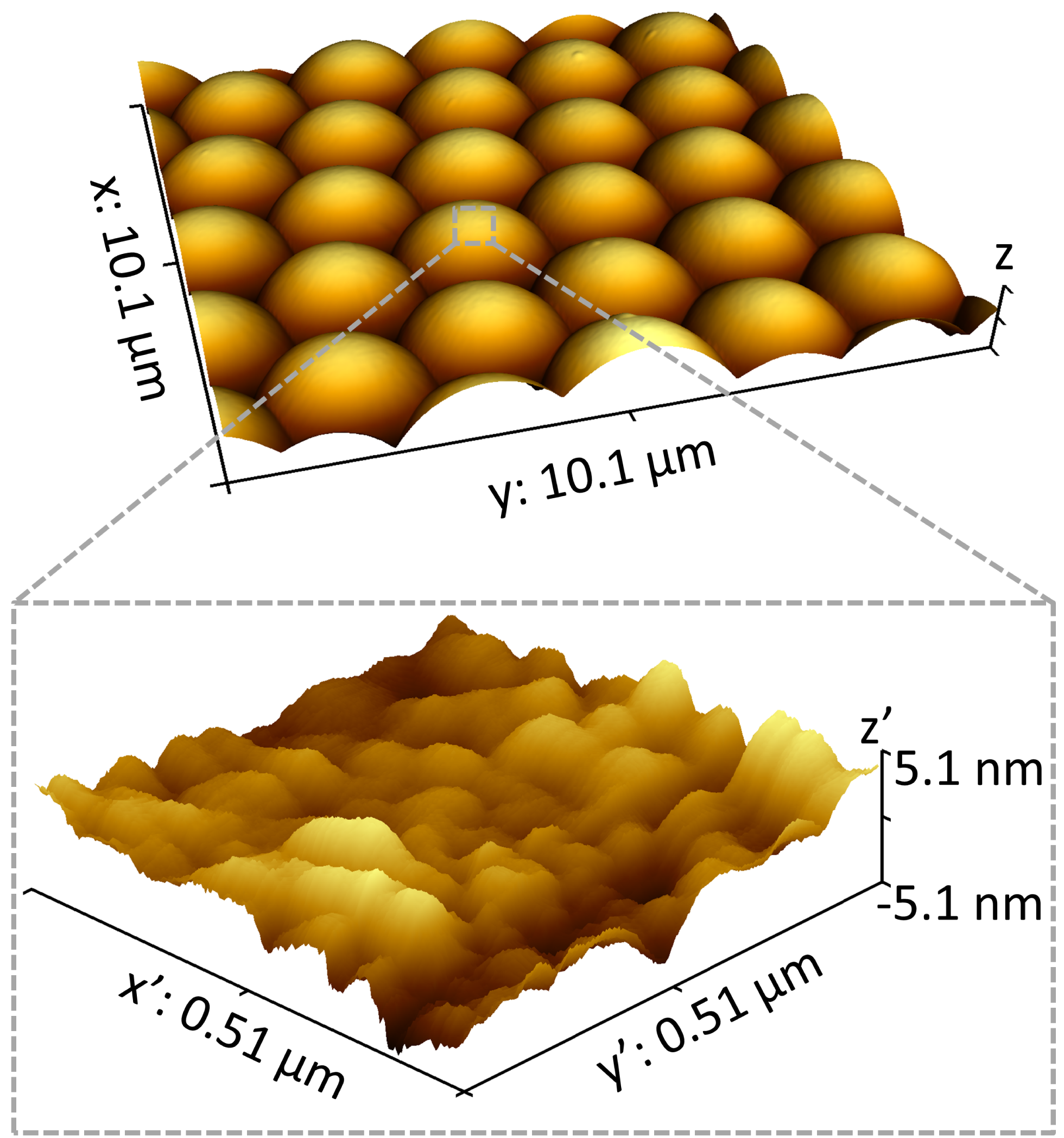}
\caption{\label{fig2} Surface roughness of the particles measured by atomic force microscopy. (Top) A scan of 10.1$\mu$m$\times$10.1$\mu$m area of the particle crystal. (Bottom) A scan of 0.51$\mu$m$\times$0.51$\mu$m area over a single particle surface. The spherical form of the particle is subtracted to obtain the roughness profile in bottom image (See SI section II for details on AFM measurements).}
\end{figure}

\begin{figure}[b]
\includegraphics[width=\linewidth]{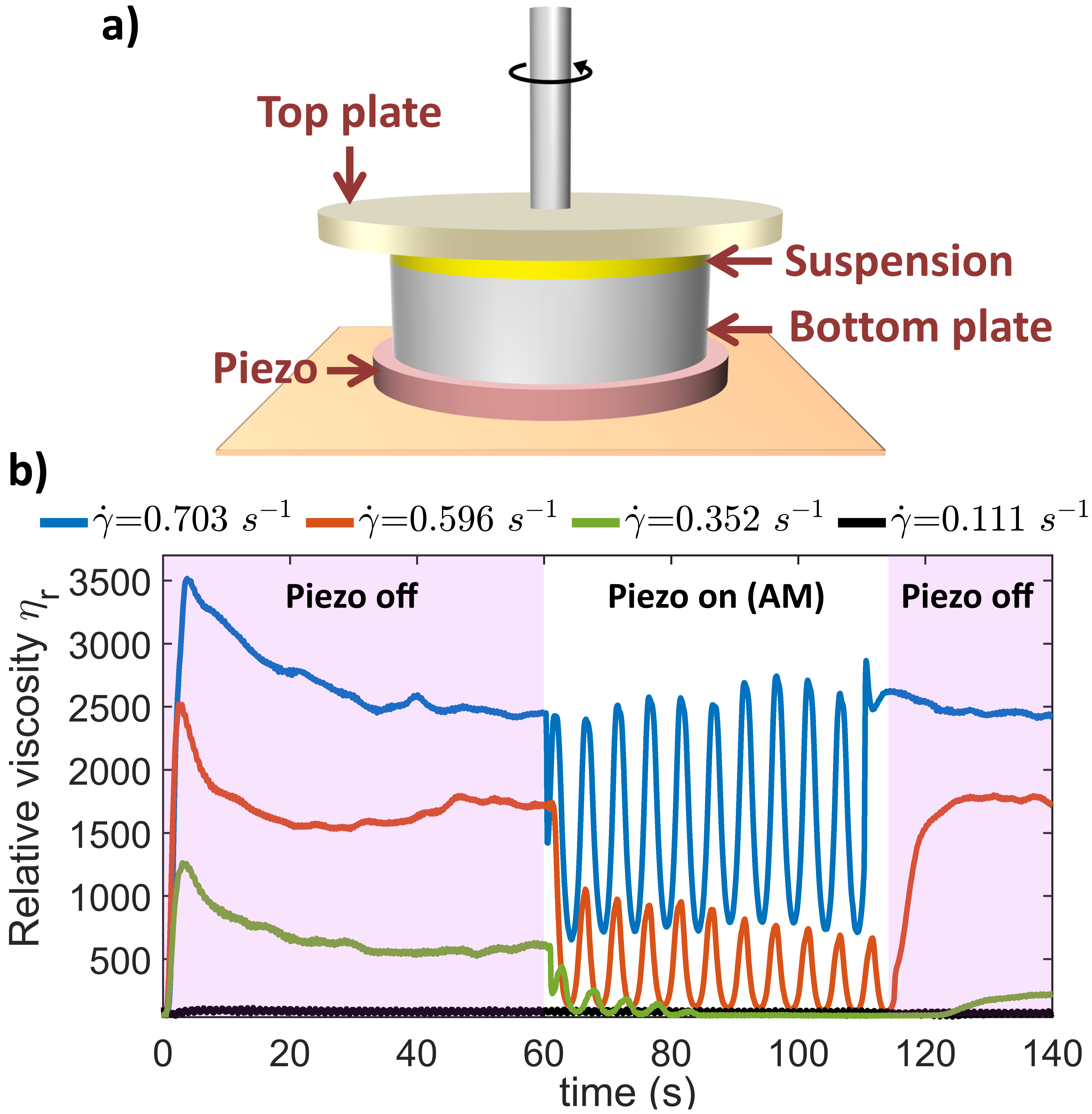}
\caption{\label{fig3} Experimental setup and AM measurements. a) The schematic of the acoustic-rheometer setup. The top plate is connected to the rheometer and the bottom plate is bonded to the piezoelectric element. The suspension is confined between the two plates. b) The instantaneous viscosity response of $\phi=0.53$ suspension to the gradient-direction perturbations at representative strain rates. The relative viscosity is defined as the ratio of the suspension viscosity to the solvent (dipropylene glycol, 0.11 Pa.s) viscosity. Each measurement is performed at a steady $\dot{\gamma}$ for 140 s in which the AM signal is turned on at time t$\sim$60~s for at least nine modulation cycles, followed by an off-period for the remaining time. Measurements for $\phi=0.50$ suspension are shown in supplementary figure~S4.}
\end{figure}

We test this idea on a dense silica colloidal suspension undergoing controlled shear and simultaneous acoustic excitation. Our suspension consists of charge-stabilized silica particles, 2 $\mu$m in diameter, in dipropylene glycol at volume fractions of $\phi=0.53$ and $\phi=0.50$. Using atomic force microscopy, we measured the average particle surface roughness to be $\sim$2~nm, which is well below the estimated dipolar oscillation amplitude (Fig.~\ref{fig2}). Our testing apparatus consists of a piezoelectric disk (APC International, Material 841) of diameter 21 mm and thickness 1.80 mm bonded via epoxy to an aluminum (6061-T6) bottom plate of diameter 19 mm and thickness 8.57 mm (Fig.~\ref{fig3}a). The acoustic perturbations are generated by exciting the piezoelectric crystal in the thickness mode at a resonance frequency $f_r =$1.15 MHz. This mode applies perturbations in the gradient direction of the primary shear flow. The bottom plate thickness is optimized for maximum energy transfer to the suspension. The piezo-plate setup is integrated with an Anton-Paar MCR702 Rheometer, and a glass top plate is used to apply the primary shear flow and measure the shear viscosity. The suspension is confined between the two plates and the gap is set to 0.64 mm (See SI Section I for calibration). Using this setup, we quantify the effects that acoustic perturbations have on the shear-thickening behavior of our suspensions.  

\begin{figure*}
\includegraphics[width=\linewidth]{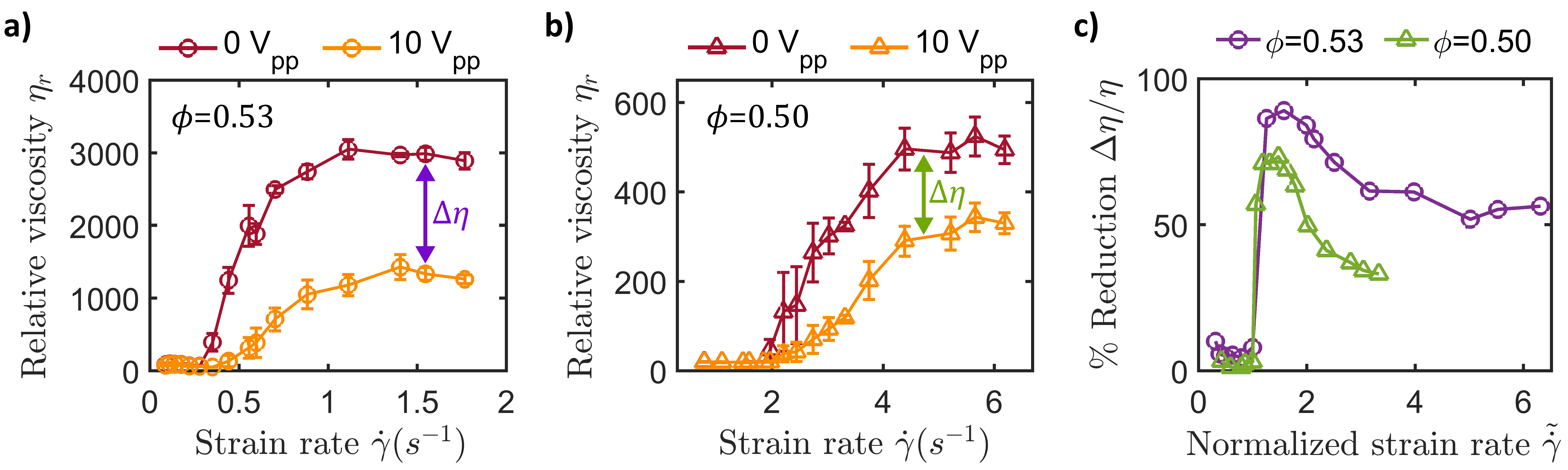}
\caption{\label{fig4} Dethickening response to the acoustic perturbations for a) $\phi=0.53$ and b) $\phi=0.50$ suspensions. Relative viscosity $\eta_r$ vs strain rate $\dot{\gamma}$ is plotted for no-perturbations (maroon curve) and 10 $V_{pp}$ perturbations (yellow curve). The viscosities are obtained from figure~\ref{fig3}b via a phase sensitive analysis that reduces temporal noise (see SI section IV for details). c) Percentage reduction in the viscosity at different normalized strain rates $\Tilde{\dot{\gamma}}$ upon application of 10 $V_{pp}$ signal in $\phi=0.53$ and $\phi=0.50$ suspensions.}
\end{figure*}

In our measurement protocol, we apply a steady shear to thicken the suspension. After the suspension reaches steady state, we add an amplitude-modulated (AM) acoustic perturbation. The voltage signal used to drive the piezoelectric element has the form $V=V_0[1+\sin{(2\pi f_m t+\Phi_0)]}\sin{(2\pi f_r t)}$, where $f_r$ is the resonance frequency, and $f_m=0.2\text{ Hz}$ is the modulation frequency (Fig.~S2). The phase $\Phi_0$ is set arbitrarily and the voltage $V_0$ is set at 2.5 V to obtain maximum peak-to-peak voltage ($V_{pp}$) of 10 V. This approach quantifies the dynamic, phase-sensitive, and power-dependent viscosity response of the suspension in a single measurement. The dynamic response probes the important time scales that govern the formation and breakup of force chain. The phase-sensitive response obtained from the controlled modulation eliminates noise in the temporal measurements of the viscosity. Finally, the power-dependent response quantifies the efficiency of this dethickening method. We perform these measurements for a range of strain rates ($\dot{\gamma}$) over which the fluid behavior varies from a Newtonian to a fully thickened state.

As hypothesized, the viscosity response of the thickened suspension depends sensitively on the acoustic perturbations (Fig.~\ref{fig3}b and Fig.~S4). For $\phi=0.53$ suspension sheared at strain rates corresponding to the thickened regime ($\dot{\gamma}=0.703 \text{ s}^{-1}$), the instantaneous relative viscosity, $\eta_r$, oscillates when the acoustic perturbations are applied. The oscillations result from the amplitude modulation of the acoustic perturbations with the greatest decrease in viscosity arising from the largest perturbation amplitude. The minima of these viscosity oscillations are still above the Newtonian viscosity, which suggests that a higher acoustic power is required to break up all the force chains. In contrast, for strain rates corresponding to the unthickened Newtonian regime ($\dot{\gamma}=0.111 \text{ s}^{-1}$) we observe no modulations in the viscosity. This difference in response is consistent with the proposed mechanism that the acoustic perturbations break the shear induced force chains responsible for thickening leaving the other suspension properties largely unchanged.

For strain rates corresponding to the transition regime between the Newtonian and fully thickened state ($\dot{\gamma}=0.596,0.352 \text{ s}^{-1}$), the acoustic perturbations are sufficient to dethicken the suspension viscosity to the value in the Newtonian regime. Interestingly, the maximum viscosity during the time when AM perturbations are applied does not recover fully to the steady state value. We interpret this response to indicate that the AM frequency is too rapid for the force chains to fully form between successive oscillations at these strain rates. This picture is supported by the fact that the viscosity recovery time when the perturbations are turned off is much longer than the AM oscillation period.     

We extract the magnitude of acoustic dethickening as a function of strain rates using a phase-sensitive analysis of the instantaneous viscosity response curves (Fig.~\ref{fig4}, see SI section IV for details). We observe that the application of the acoustic perturbations decreases the viscosity substantially in the regime where the suspension thickens. This response is sensitive to the strain rate, with the largest decrease occurring in the transition regime (Fig.~\ref{fig4}a,b). We quantify this response by plotting the $\%$Reduction in viscosity versus $\Tilde{\dot{\gamma}}$, the strain rate normalized by the strain rate at the onset of thickening  (Fig.~\ref{fig4}c). We find negligible decrease in the viscosity in the Newtonian regime ($\Tilde{\dot{\gamma}}<1$), in which the force chains are mostly absent. We find the highest reduction in the transition region ($1<\Tilde{\dot{\gamma}}<2$), in which the applied acoustic perturbations are sufficient to break up the majority of the force chains. This decrease in viscosity to nearly the Newtonian value effectively shifts the onset strain rate for thickening. Finally, we find that the $\%$Reduction decreases and plateaus in the fully thickened regime ($2<\Tilde{\dot{\gamma}}$). This plateau is consistent with literature predictions that the force chain network saturates in the thickened regime \cite{MariandSeto_ForceChains} and with the idea that the acoustic perturbations are only breaking up a fraction of this network at this power. The trends at each volume fraction are similar, but dethickening is lower in  $\phi=0.50$ suspensions than in $\phi=0.53$ suspensions. This difference in effect magnitude is consistent with the results from boundary oscillation simulations \cite{Ness_CrossShear}. Collectively, these data suggest that a further increase in power would shift the onset of the thickening to higher strain rates or stress scales \cite{Wyart_DSTFriction}, and increase the dethickening in the thickened regime.

\begin{figure}[t]
\includegraphics[width=\linewidth]{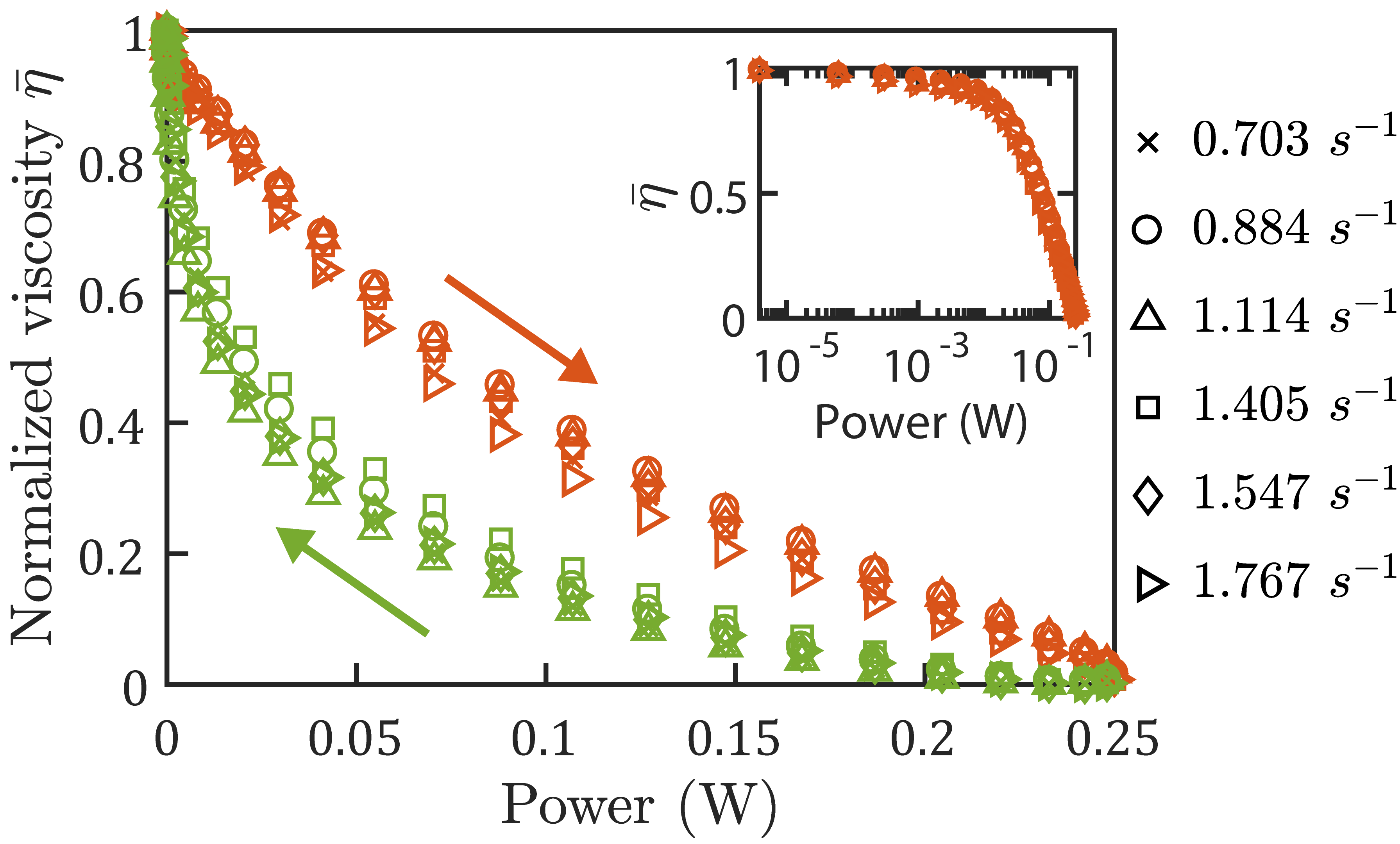}
\caption{\label{fig5}  Evolution of viscosity with input power at different strain rates (markers) for $\phi=0.53$ suspensions. Normalized viscosity $\bar{\eta}$ is calculated from the Fourier analysis of the data in figure~\ref{fig3}b (See SI section IV). Orange markers show the decrease in viscosity when the power is ramped up and green markers show the subsequent increase in viscosity when the power is ramped down. A power analysis confirms that thermal contributions to the observed dethickening response are negligible (See SI Section VI). (Inset) Evolution of normalized viscosity with the ramp up of power (orange markers) on a semi-log plot. Power-dependent response for $\phi=0.50$ suspensions is shown in supplementary figure S5.}
\end{figure}

 Using the AM protocol, we determine the power-dependent response of the suspension and observe a logarithmic decrease in the viscosity above a threshold power (Fig.~\ref{fig5}). We define a normalized viscosity $\bar{\eta}=(\eta-\eta_{min})/(\eta_{max}-\eta_{min})$, where $\eta_{max}$ and $\eta_{min}$ are the maximum and minimum viscosities at given $\dot{\gamma}$ during a modulation period. We plot $\bar{\eta}$ versus power for strain rates in which the turnover time of force chains is substantially faster than the time period of the AM signal (e.g. blue curve in figure~\ref{fig3}b). Strikingly, we find a threshold-like behaviour at very low powers (inset Fig. 5), where the viscosity decrease is very slow and negligible up to powers $\sim$5~mW (corresponding to $V_{pp}=1.4$ V). This trend is consistent with our mechanistic hypothesis that a minimum particle displacement is required to break force chains. Beyond this threshold, the viscosity initially decreases rapidly at lower powers up to $\sim$ 0.08~W  (orange curve in Fig. 5), which suggests that a significant number of force chains are broken even at a small input power. At higher power, the viscosity decreases more slowly. This non-linear evolution suggests that although a higher power is required to completely eliminate the thickening effects of force chains, a large fraction of dethickening can be achieved by just small input power.
 
 As the power is ramped down (green curve), we observe a hysteretic response indicating that while a larger power is needed to break up the force chains, a significantly lower power is needed to maintain this disruption during this period. These results suggest that the suspension retains some memory of the microstructure within each AM cycle even though the modulation time period is large enough for the force chains to fully recover. Our findings are consistent with the simulations of orthogonal boundary oscillations that indicate a dethickened viscosity can be maintained using pulsed perturbations \cite{Ness_CrossShear}. Our ability to precisely measure these hysteresis loops shows that this amplitude modulation technique can determine the relevant time scales to form and break force chains. This understanding can then be leveraged to optimize strategies for achieving dethickening with minimal amount of power.

 For the current implementation of these acoustic modulations, the power required to dethicken the suspension is over an order of magnitude higher than the power required for shearing the thickened suspension. Further studies, aimed at improving the coupling of acoustic energy to the suspension and applying perturbation intermittently, may change this balance. Moreover, the particle displacements that are driving these changes in the viscosity maybe complex in the dense suspensions because of the acoustic scattering from nearby particles, bulk acoustic flows, and steric constraints. Measurements or simulations of the exact particle displacements are required to confirm their hypothesized mechanistic link with the particle roughness. This understanding of the suspension microstructure upon acoustic excitation and development of strategies for enhancing disruptions should also enable more efficient control of the suspension viscosity.

 Even in its current form, however, our method has paramount advantages in the applications where the goal is to increase the flow rate, unclog a system, or control the viscosity, without energy concerns. Such applications include high-throughput processing of dense suspensions, avoiding jamming in narrow conduits, 3D printing, and designing of smart materials. In each of these cases, the perturbations can be applied by simply bonding a piezoelectric element to a fixed surface, which makes this method easy to integrate with the existing practical systems without modifying their geometry. Furthermore, acoustic perturbations can tune thickening in these applications in multiple modes, thus providing flexibility in implementation (See SI Section VII for data on acoustic perturbations in the vorticity direction). Overall, our method has laid a strong foundation to robustly design smart transport systems that handle shear-thickening fluids.

We thank Anton Paar for use of the MCR 702 rheometer through their VIP academic research program, Abhishek Shetty for help with the rheometer, the Kirby group and the Cohen group for their insightful suggestions, Prof. Wolfgang Sachse at Cornell for lending equipment and useful discussions, and the Amit Lal group at Cornell for providing the impedance analyzer. This work is supported by NSF CBET award numbers 1804963, 1232666, and 1509308, partially supported by the Center on the Physics of Cancer Metabolism through Award Number 1U54CA210184-01, and performed in part at the Cornell NanoScale Facility, an NNCI member supported by NSF Grant NNCI-1542081. PS and MR contributed equally to this work.

\bibliography{references}
\bibliographystyle{apsrev4-1}
\end{document}


\title{Using Acoustic Perturbations to Dynamically Tune Shear Thickening in Colloidal Suspensions}

\author{Prateek Sehgal}
\affiliation{Sibley School of Mechanical and Aerospace Engineering, Cornell University, Ithaca, New York 14853, USA}
\author{Meera Ramaswamy}
\affiliation{Department of Physics, Cornell University, Ithaca, New York 14853, USA}
\author{Itai Cohen}
\affiliation{Department of Physics, Cornell University, Ithaca, New York 14853, USA}
\altaffiliation[Also at ]{Department of Medicine, Division of Hematology and Medical Oncology, Weill--Cornell Medicine, New York, New York 10021,USA}
\author{Brian J. Kirby}
\affiliation{Sibley School of Mechanical and Aerospace Engineering, Cornell University, Ithaca, New York 14853, USA}
\affiliation{Department of Medicine, Division of Hematology and Medical Oncology, Weill--Cornell Medicine, New York, New York 10021,USA}

\date{\today}

\pacs{Valid PACS appear here}

\maketitle
\renewcommand{\thefigure}{S\arabic{figure}}

\section{Calibration of the setup}
The apparatus consists of an Anton Paar MCR 702 rhometer where the bottom plate is replaced with a custom-made aluminum plate attached to a piezoelectric crystal. We recalibrated the system with the custom plate to ensure correct measurements of viscosities and strain rates. To calibrate the average strain rate applied to the sample, we scale the strain rate set in the rheometer by the ratio of the plate diameters $\dot{\gamma} = \dot{\gamma}_\textrm{set} \times 19/43$. To calibrate the viscosity, we measure a viscosity standard (N4000 - ISO 17025) in both the standard parallel-plate configuration and our apparatus comprising the modified bottom plate with piezo. The ratio of the two viscosities is used to scale all the measured viscosities. 

\begin{figure}[htbp!]
\includegraphics[width=\linewidth]{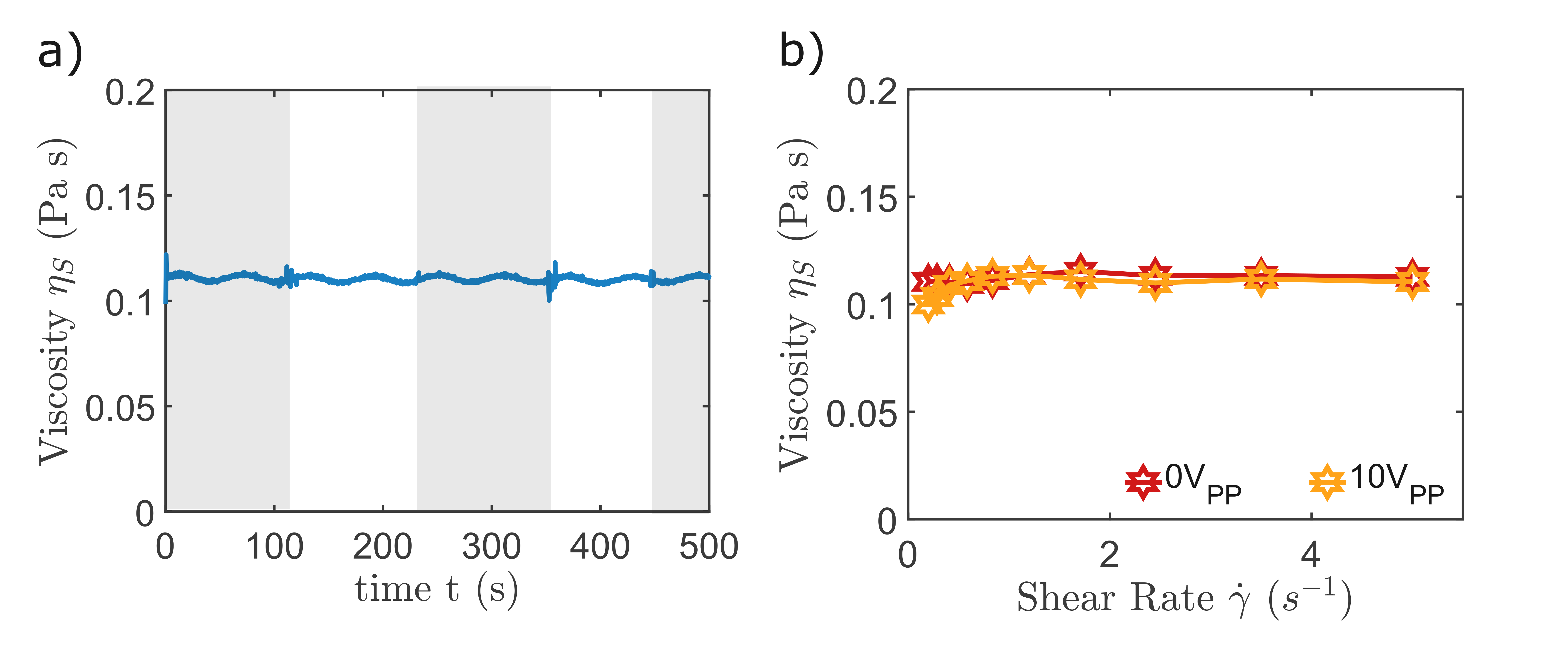}
\caption{\label{figS1} The response of dipropylene glycol to acoustic perturbations. a) Plot of the viscosity versus time as the AM acoustic perturbations are turned on (white regions) and off (grey regions). We observe negligible difference in the viscosity between each condition.  b) Viscosity versus strain rate for piezo off (red curve) and piezo continuously on at a fixed amplitude corresponding to the peak-to-peak voltage of $10$ V (yellow curve). The difference between the two curves is negligible.}
\end{figure}

We test the viscosity response of the Newtonian solvent used in the experiments, dipropylene glycol, to the acoustic perturbations. We turn on the AM acoustic perturbations between $t = 110s - 230s$ and between $t = 350s - 450s$ (white bands in Fig.~\ref{figS1}a). We see no observable change in the viscosity. In addition, we performed shear sweeps over a range of shear rates with the piezo continuously on at a peak-to-peak voltage of $10$ V (yellow curve) and with the piezo off (red curve) as shown in figure~\ref{figS1}b. We observe negligible difference between the two measurements, confirming that the dethickening in acoustically perturbed colloidal suspensions is not associated with changes in the solvent viscosity.

\FloatBarrier

\section{Atomic Force Microscopy}
The atomic force microscopy (AFM) measurements are performed in Bruker Dimension Icon using Arrow-NC cantilever tip (from Nanoandmore USA, shape - arrow, radius $<$ 10nm). To prepare the sample, the powdered silica particles are dissolved in ethanol, and then dried on the silicon substrate. The AFM measurements are performed in tapping mode because the silica particles are loosely attached to the silicon substrate. The surface profile data from AFM is post processed in Gwyddion 2.52 to obtain the roughness. Briefly, the horizontal strokes are filtered, a spherical shape is fitted to obtain the form of the particles, and the spherical form is subtracted to obtain the flattened roughness profile in figure 2.

\section{Amplitude modulation of the voltage signal}
\begin{figure}[!htbp]
\includegraphics[width=0.85\linewidth]{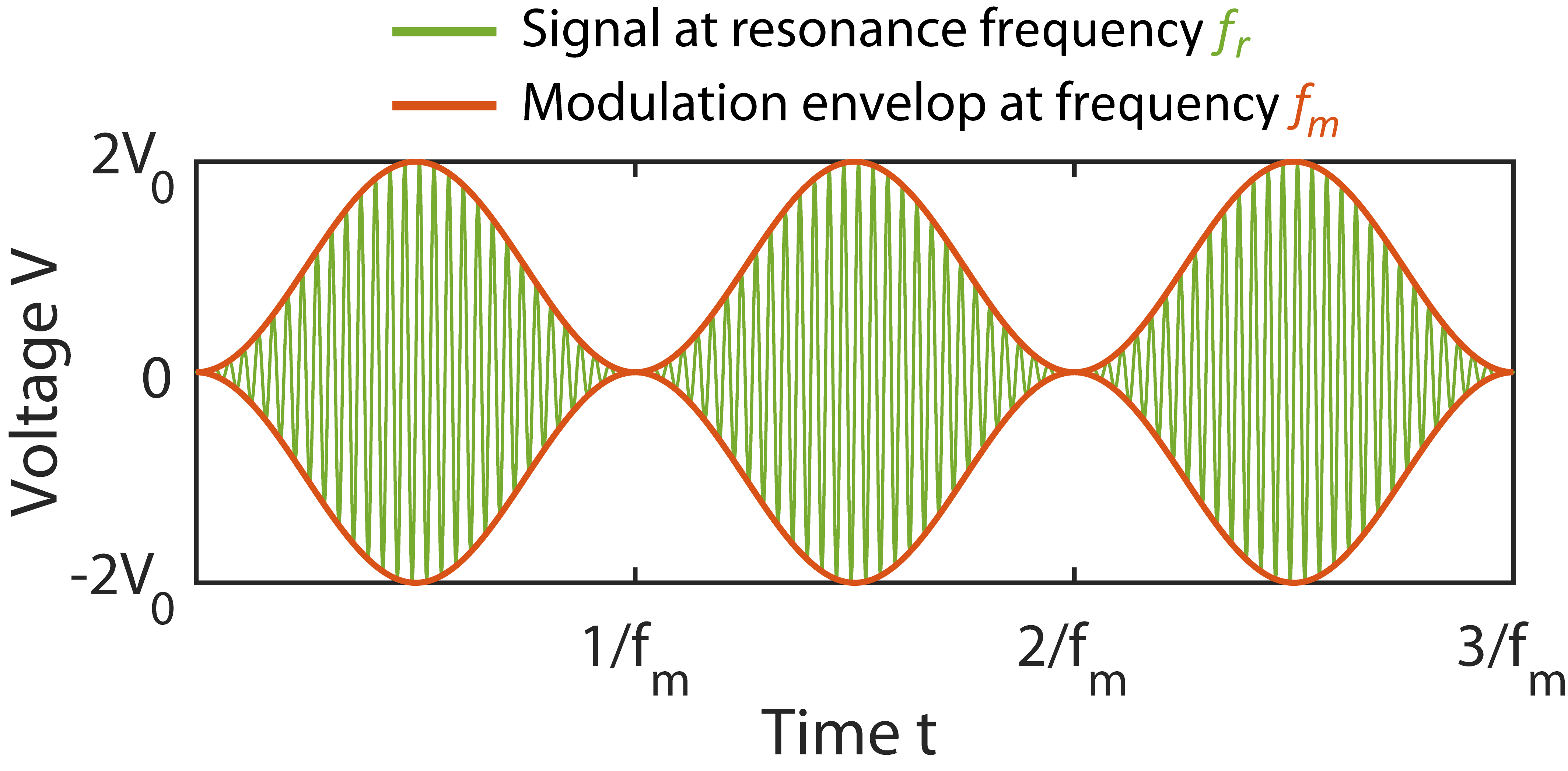}
\caption{\label{figS2}  Amplitude-modulated voltage signal. The amplitude of the voltage signal at resonance frequency $f_r$ (green curve) is modulated at the modulation frequency $f_m$ (orange envelop). The graph shows three cycles of the modulation with a maximum amplitude of $2V_0$ corresponding to $10 V_\textrm{PP}$.}
\end{figure}

We use an amplitude modulated signal $V=V_0[1+\sin{(2\pi f_m t+\Phi_0)]}\sin{(2\pi f_r t)}$ to measure the suspension viscosity dynamically as a function of power and use the phase-sensitive modality to reduce measurement noise.  An example of such an amplitude-modulated signal is shown in figure~\ref{figS2}. Here, we use a resonance frequency, $f_r$, of 1.15 MHz or 107 kHz depending on the excitation mode of the piezo and a (much slower) modulation frequency $f_m = 0.2$ Hz. We apply the modulation with a minimum amplitude of $0$ V and a maximum amplitude of $2V_0$ that corresponds to a peak-to-peak voltage of $10$ V. Thus, we expect the dethickening to be maximum when the amplitude is $2V_0$ ($V_{pp}=10$ V) and minimum when the amplitude is $0$ V.

\FloatBarrier
\section{Measurement of Acoustic Dethickening}

We use the instantaneous viscosity response from AM measurements to extract the magnitude of acoustic dethickening at different strain rates and voltages. From preliminary steady state experiments, we find that the timescales for force chain break up is of the order $100$ ms. Thus, we expect that the ramp up period of 2.5s of the modulations is sufficient to capture the quasi-steady dethickening behaviour of the suspension. To quantify the magnitude of acoustic dethickening from the instantaneous viscosity response, we extract the viscosities at 0 $V_{pp}$ and 10 $V_{pp}$ voltages. Here, we choose methods that minimize the noise in the measurements as detailed in the following paragraphs. 

\begin{figure}[!htbp]
\includegraphics[width=0.85\linewidth]{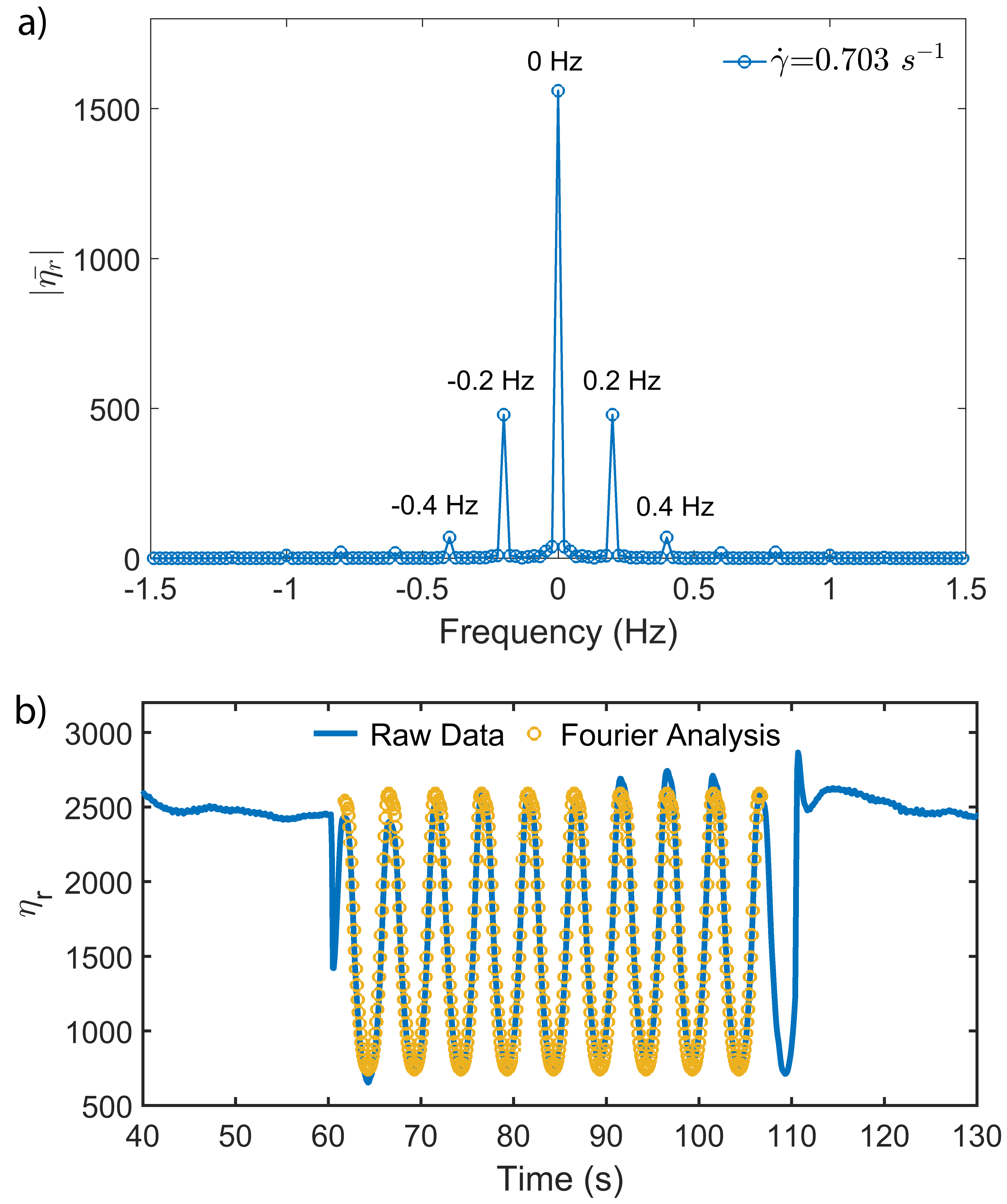}
\caption{\label{figS3}  Phase-sensitive analysis. a) Fourier transform of the instantaneous viscosity response at $\dot{\gamma}=0.703 \text{ s}^{-1}$ from figure 3b. The absolute magnitude of the complex relative viscosity $\bar{\eta_{r}}$, obtained from the Fourier transform, is plotted in the frequency space. b) The instantaneous viscosity response is re-constructed (yellow markers) from the Fourier transform of raw data by extracting the signal at 0 Hz (DC response) and the signal at the driving frequency ($\pm$ 0.2 Hz) and its higher harmonics ($\pm$ 0.4 Hz, $\pm$ 0.6 Hz, ...).}
\end{figure}

For strain rates in the fully thickened regime (blue curve in Fig. 3b) and the Newtonian regime (black curve in Fig. 3b), we use phase-sensitive analysis to reduce noise and measure the averaged dethickening response. Briefly, we take Fourier transform of the instantaneous viscosity data for 9 AM cycles as shown in figure~\ref{figS3}a. From the Fourier space, we extract the viscosity signal at the driving frequency ($f_m = 0.2$ Hz) and its higher harmonics to reconstruct the instantaneous viscosity response with reduced temporal noise (Fig.~\ref{figS3}b). We use the maximum and minimum of this reconstructed viscosity response to obtain the relative viscosity corresponding to 0 $V_{pp}$ and 10 $V_{pp}$ voltages, respectively, in figure 4. We also use this reconstructed viscosity response to calculate the normalized viscosity $\bar{\eta}$ and plot it against the input power in figure 5. The maximum ($\eta_{max}$) and minimum ($\eta_{min}$) viscosities described in the definition of $\bar{\eta}$ are the viscosities corresponding to 0 $V_{pp}$ and 10 $V_{pp}$ voltages in figure 4. 

For strain rates in the transitioning regime (orange and green curve in Fig. 3b), where the maximum viscosity during AM does not recover fully to the steady state piezo-off value, we can no longer use the Fourier analysis in AM duration to estimate the viscosity at 0 $V_{pp}$ voltage. Instead, the viscosity at 0 $V_{pp}$ is obtained by averaging $\eta_r$ from 50 s to 60 s in the steady state piezo-off region. The viscosity at 10 $V_{pp}$ signal is obtained by averaging the minima during modulations in piezo-on region.

\FloatBarrier

\section{Tunable shear thickening response for 0.50 volume fraction suspensions}
\begin{figure}[!tbp]
\includegraphics[width=0.85\linewidth]{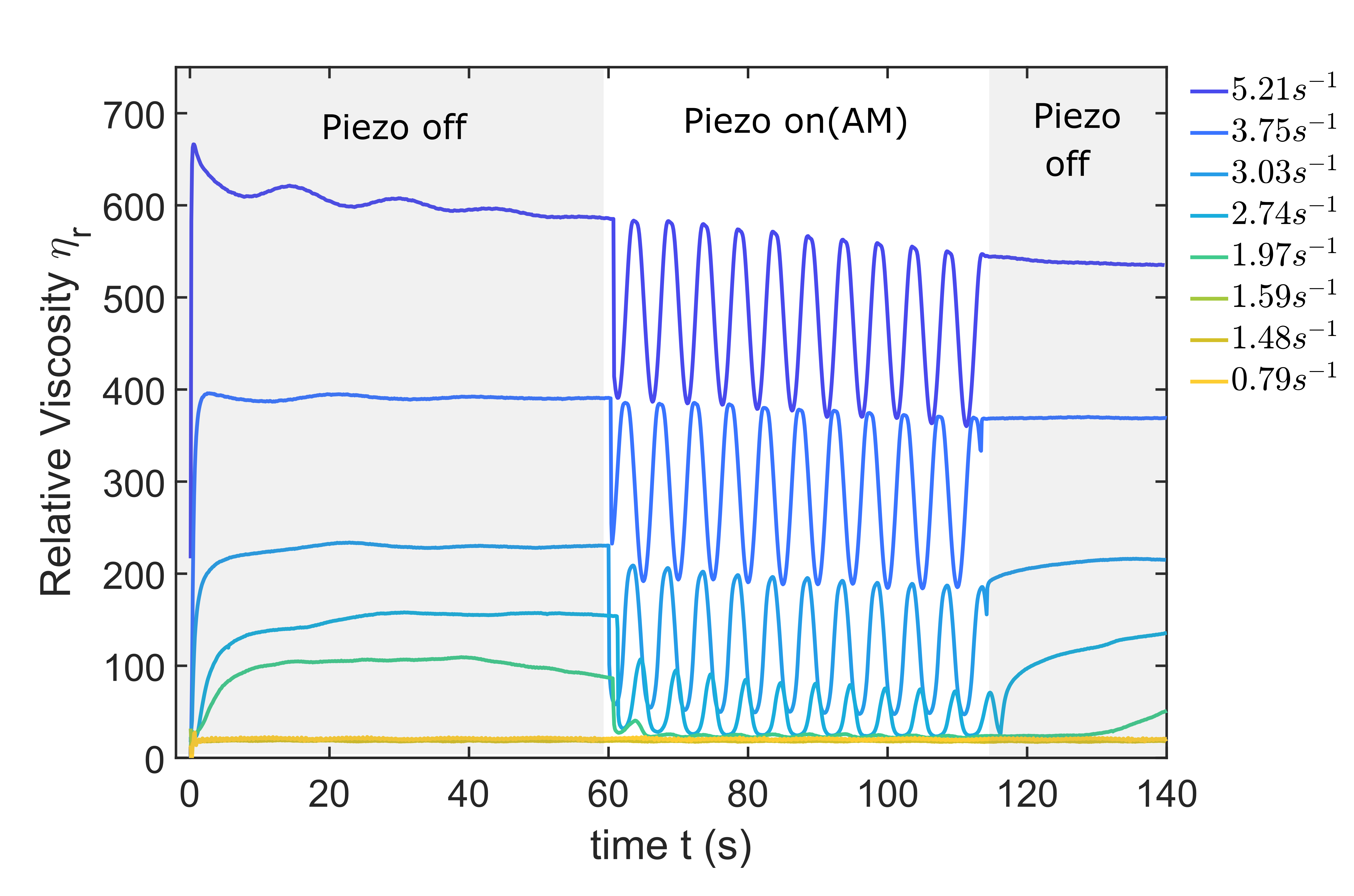}
\caption{\label{figS4}  Dynamic tuning of shear thickening in $\phi=0.50$ suspensions. The figure shows the instantaneous viscosity response of the suspension to the gradient-direction perturbations at representative strain rates. Each measurement is performed at a steady $\dot{\gamma}$ for 140 s in which the AM signal is turned on at time  t$\sim$60~s for at least nine modulation cycles, followed by an off-period for the remaining time.}
\end{figure}

The instantaneous viscosity response of $\phi = 0.50$ suspension to the gradient-direction perturbations is shown in figure~\ref{figS4}. The suspension is allowed to reach equilibrium at each strain rate, and then AM acoustic perturbations are turned on for atleast 9 AM cycles, followed by an off-period for remaining time. The $\phi=0.50$ suspension behaves similarly to the $\phi = 0.53$ suspension, where we find modulation of viscosity in the fully thickened regime and the transition regime and no modulation in the Newtonian regime. For $\dot{\gamma} = 5.21 s^{-1}$ and $\dot{\gamma} = 3.75s^{-1}$ corresponding to the fully thickened state, the rate of formation of force chains is faster than the AM time period, and thus the maximum viscosity recovers fully to the steady state value after each AM cycle. For $\dot{\gamma} = 2.74 s^{-1}$ and $\dot{\gamma} = 1.97 s^{-1}$ corresponding to the transition regime, the rate of formation of force chains is slower than the AM period, and therefore the maximum viscosity in each AM cycle is smaller than the steady-state value. For $\dot{\gamma} = 1.48 s^{-1}$ and $\dot{\gamma} = 0.79 s^{-1}$, the suspension is Newtonian, and we observe no change in viscosity upon application of the acoustic perturbations. Collectively, these data recapitulate the results in the main manuscript for this lower volume fraction.

\begin{figure}[!htbp]
\includegraphics[width=0.9\linewidth]{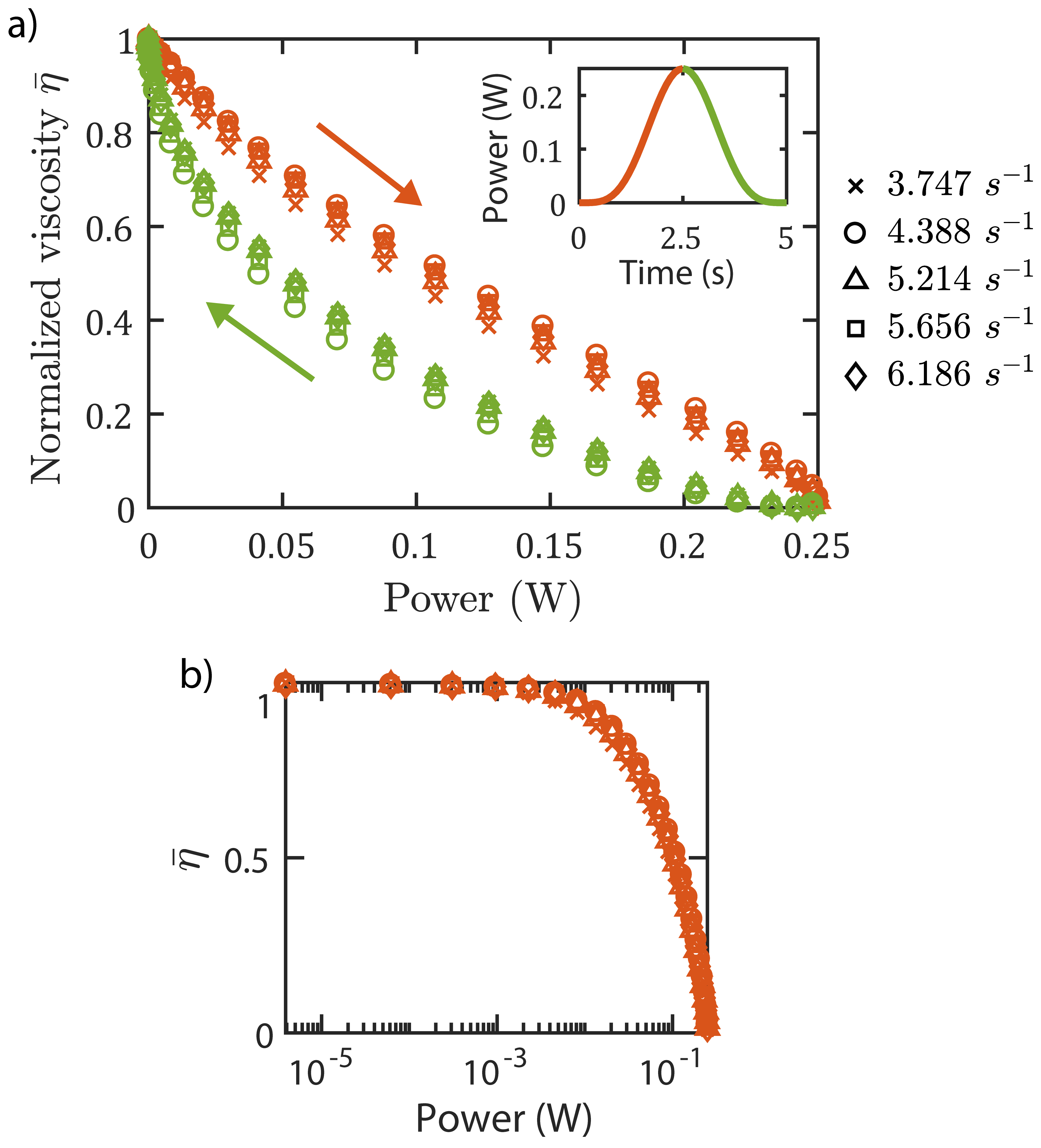}
\caption{\label{figS5}  a) Evolution of viscosity with input power at different strain rates (markers) for $\phi=0.50$ suspensions. The perturbations are applied in the gradient direction. The normalized viscosity $\bar{\eta}$ is calculated from the Fourier analysis of the data in figure~\ref{figS4}. Orange markers show the decrease in viscosity when the power is ramped up and green markers show the subsequent increase in viscosity when the power is ramped down. (Inset) The variation of power in one AM cycle, where the orange curve denotes increasing power and the green curve denotes decreasing power. b) Evolution of normalized viscosity with the ramp up of power (orange markers in (a)) on a semi-log plot.}
\end{figure}

The evolution of viscosity with input power for $\phi = 0.50$ suspensions is shown in figure ~\ref{figS5}. We see a similar threshold-like behavior at very low powers, non linear response, and hysteresis curve as seen for $\phi = 0.53$ suspensions (Fig.~5). The slight variation in the shape of the curve in comparison to the larger volume fraction likely arises from the differences in the force chain network at different volume fractions. 

\FloatBarrier

\FloatBarrier

\section{Thermal Analysis}
\begin{figure}[!tbp]
\includegraphics[width=\linewidth]{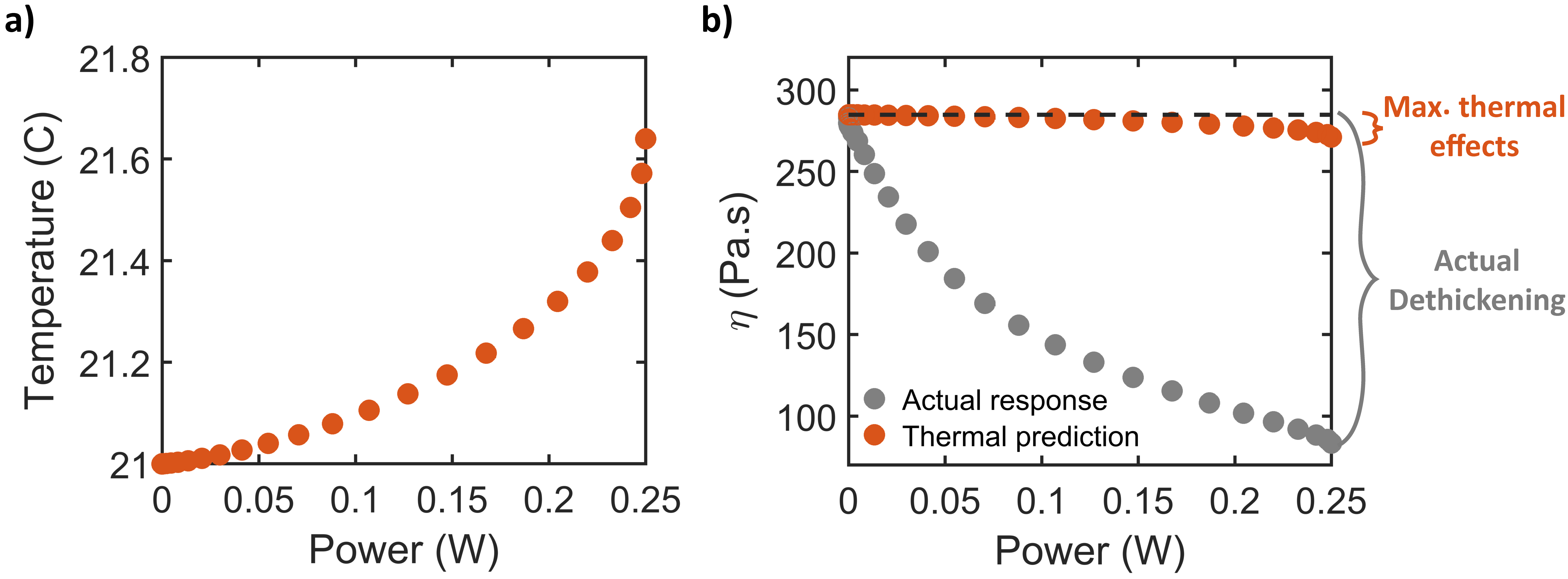}
\caption{\label{figS6}  Thermal Analysis for $\phi=0.53$ suspensions. a) Estimated temperature of the suspension versus input power assuming that all the input energy is thermally dissipated. b) (Orange curve) Variation in the viscosity with input power due to thermal dissipation. (Grey curve) Actual viscosity response obtained in our experiments from gradient-direction perturbations.}
\end{figure}

We perform a thermal analysis of the system to confirm that the observed dethickening response is not related to any thermal absorption of acoustic energy in the suspension. We estimate the viscosity response that would occur if all the input acoustic power is thermally dissipated in the suspension to increase its temperature and if there is no cooling of the system. These limiting conditions estimates a maximum decrease in viscosity that can occur due to thermal absorption of the input power. For simplicity, we assume a lumped capacitance model for these calculations which means that the temperature is uniform in the suspension. With these approximations, the governing equation for energy balance is $mc_{p}dT/dt=P_\textrm{in}$, where $m$ is the mass, $c_p$ is the specific heat capacity, and $T$ is the temperature of the suspension. The input power $P_\textrm{in}$ is given by $P_\textrm{in}={V_0}^2[1-\cos{(2\pi f_m t)}]^2/2R$, where R is the electrical resistance of the resonator. The energy balance equation is integrated from the initial condition $T=T_{a}$ at $t=0$ to obtain the instantaneous temperature variation of the suspension as $T=T_{a}+\frac{{V_0}^2}{2Rmc_p}[1.5t-\frac{\sin{(2\pi f_m t)}}{{\pi}f}+\frac{\sin{(4\pi f_m t)}}{8{\pi}f}]$, where $T_a$ is the ambient temperature. From this equation, we plot the estimated upper bound on the increase in temperature with the power over a half period of the AM cycle (Fig.~\ref{figS6}a). At the maximum input power, the temperature increases by less than a degree Celsius. To obtain the resultant viscosity change due to the increase in temperature, we measure the viscosity in a thermally controlled setup of the rheometer in the thickened regime.The viscosity changes linearly with temperature over the small temperature range of our interest, and therefore we assume that $d\eta/dT$ is a constant. Combining these measurements with the governing equation for the instantaneous temperature, we obtain the instantaneous viscosity response due to thermal dissipation as $\eta_{T}=\eta_{0}+\frac{{V_0}^2d\eta/dT}{2Rmc_p}[1.5t-\frac{\sin{(2\pi f_m t)}}{{\pi}f}+\frac{\sin{(4\pi f_m t)}}{8{\pi}f}]$. Here, $\eta_{0}$ is the ambient viscosity of the suspension at $t=0$. Using this equation, we plot the contribution of the thermal dissipation to the power dependence of the viscosity, and compare this contribution to the measured dethickening response in our experiments (Fig.~\ref{figS6}b). Clearly, the thermal decrease in viscosity with the input power (orange) is negligible relative to the experimentally observed decrease in viscosity (grey). Notably, this thermal decrease in viscosity is an over estimation of the thermal effects because, for these calculations, we assume that \emph{all} the input power is thermally dissipated in the suspension and there is no cooling of the system. These results lead us to conclude that the dethickening response measured in our system has negligible contributions from the thermal effects. 

\FloatBarrier

\section{Tuning Shear Thickening by Perturbations in Vorticity Direction}

\begin{figure}[!tbp]
\includegraphics[width=\linewidth]{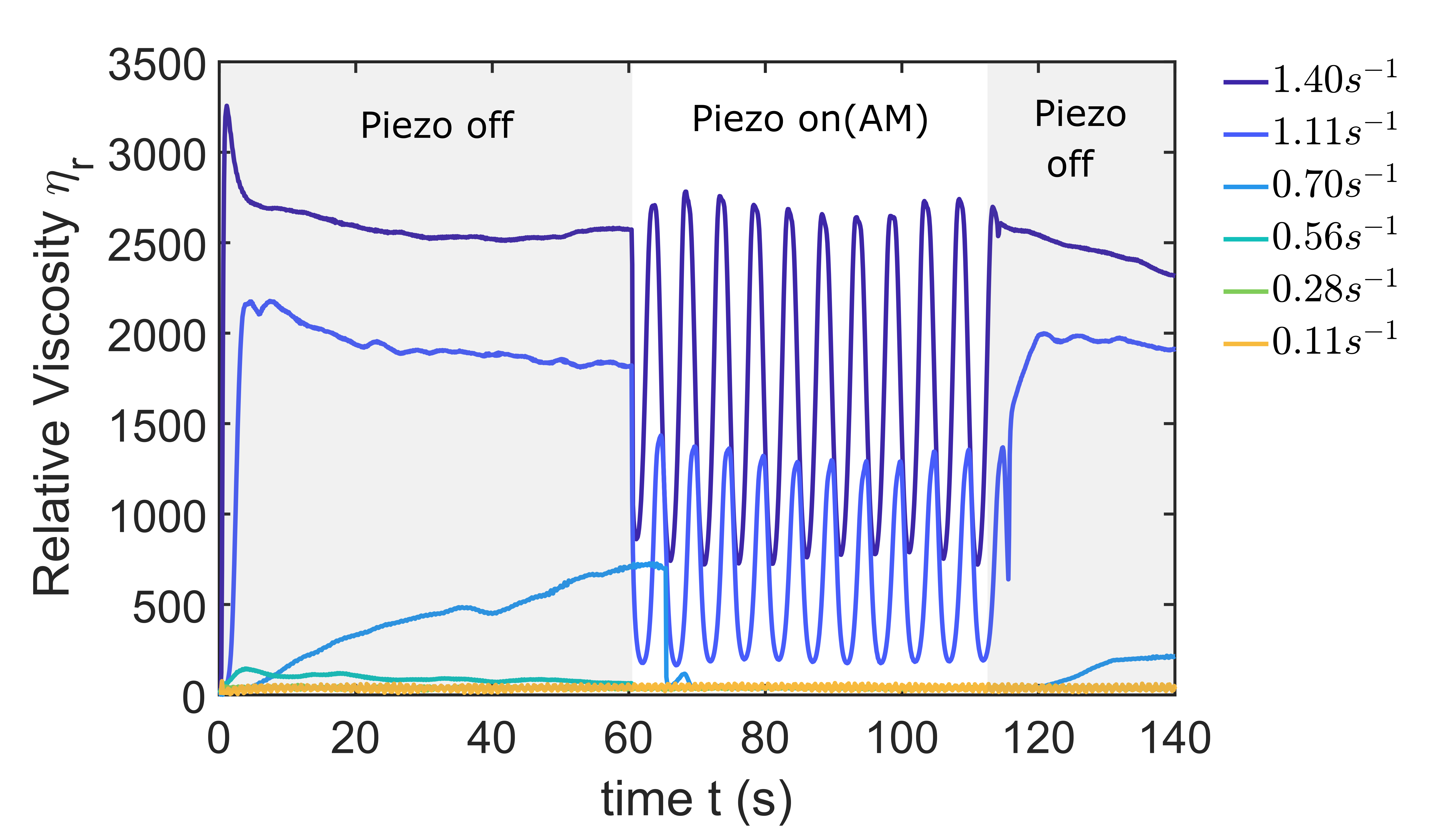}
\caption{\label{figS7}  Dynamic tuning of shear thickening by vorticity-direction perturbations. The figure shows the instantaneous viscosity response of $\phi=0.53$ suspension to the vorticity-direction perturbations at representative strain rates. Each measurement is performed at a steady $\dot{\gamma}$ for 140 s in which the AM signal is turned on at time  t$\sim$60~s for at least nine modulation cycles, followed by an off-period for the remaining time.}
\end{figure}

\begin{figure}[!htbp]
\includegraphics[width=\linewidth]{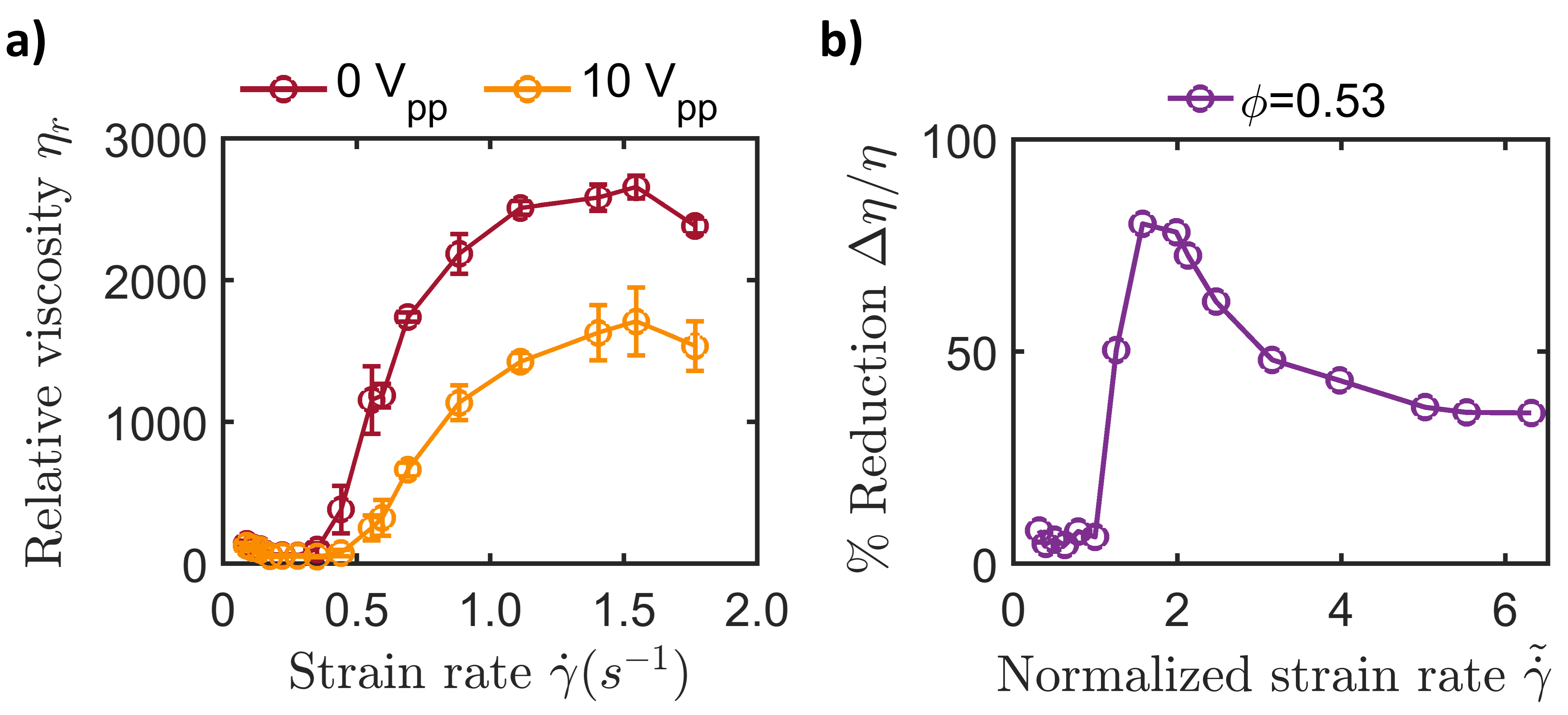}
\caption{\label{figS8} Dethickening response to the vorticity-direction perturbations for $\phi=0.53$ suspensions. a) The relative viscosity $\eta_r$ vs strain rate $\dot{\gamma}$ without acoustic perturbations (maroon curve) and with 10 $V_{pp}$ acoustic perturbations (yellow curve). The viscosities are obtained from figure~\ref{figS7} through Fourier analysis to reduce temporal noise. b) Percentage reduction in viscosity at different normalized strain rates $\Tilde{\dot{\gamma}}$ upon the application of 10 $V_{pp}$ signal.}
\end{figure}

\begin{figure}[!htbp]
\includegraphics[width=0.9\linewidth]{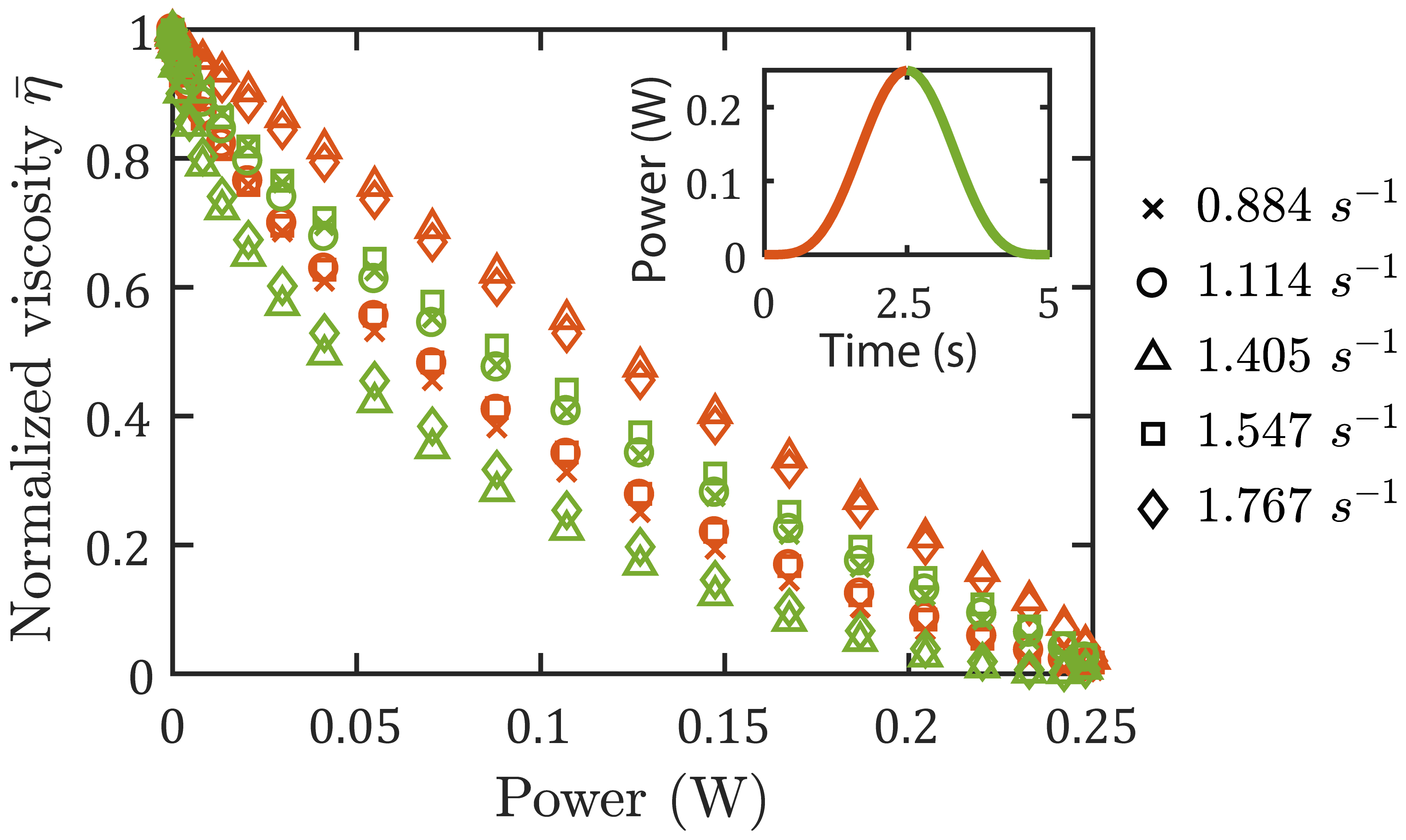}
\caption{\label{figS9}  Evolution of viscosity with input power at different strain rates (markers) for vorticity-direction perturbations applied to $\phi=0.53$ suspensions. The normalized viscosity $\bar{\eta}$ is calculated from the Fourier analysis of the data in figure~\ref{figS7}. Orange markers show the decrease in viscosity when the power is ramped up and green markers show the subsequent increase in viscosity when the power is ramped down. (Inset) The variation of power in one AM cycle, where the orange curve denotes increasing power and the green curve denotes decreasing power.}
\end{figure}

To explore different modalities for tuning shear thickening, we apply acoustic perturbations in the vorticity direction of the primary shear flow, which is analogous to the oscillation direction imposed in \cite{Lin_dethick}. We generate these perturbations in $\phi=0.53$ suspensions by driving the piezoelectric element in the radial mode at the resonance frequency $f_r=107$ kHz. We find a dynamically tunable dethickening response similar to that of gradient-direction perturbations (Fig.~\ref{figS7},~\ref{figS8}) because each of these directions is orthogonal to the primary shear flow and the perturbations are applied faster (MHz/kHz) than the frequency of force-chain formation ($\sim\dot{\gamma}$). These fast perturbations are critical to breaking the force chains responsible for shear thickening. We observe a slightly lower dethickening in the radial mode than in thickness mode, because our resonator is primarily optimized for the maximum energy transfer in the thickness mode. The dependence of $\bar{\eta}$ on power for the vorticity-direction perturbations shows a  non-linear trend (Fig.~\ref{figS9}) similar to that shown in figure~5 for the gradient-direction perturbations. The curves at different strain rates, however, do not collapse to a single curve as that in figure~5 and figure~\ref{figS5}. Importantly, these results show that the tunable viscosity can be achieved by applying fast perturbations in either of the two directions that are orthogonal to the primary shear flow.

\FloatBarrier

\FloatBarrier

\FloatBarrier
\bibliography{references.bib}
\bibliographystyle{apsrev4-1.bst}